\documentclass[11pt,letterpaper,english,rmp,amsmath,amssymb,superscriptaddress]{revtex4}
\usepackage[T1]{fontenc}
\usepackage[latin1]{inputenc}
\usepackage{color}
\usepackage{babel}
\usepackage{graphicx}
\usepackage[unicode=true,bookmarks=false,breaklinks=false,pdfborder={0 0 1},colorlinks=false]{hyperref}
\usepackage{breakurl}

\usepackage{float}
\usepackage{eurosym}
\usepackage{filemod}
\usepackage{soul}
\usepackage{dcolumn}
\usepackage{bm}
\usepackage{rotating}
\usepackage{psfrag}
\usepackage{longtable}
\usepackage{ulem}
\makeatletter

\def\p2{\langle p^2 \rangle}

\def\kbar{{\mathchar'26\mkern-9mu k}}
\def\rme{\mathrm{e}}
\def\rmi{\mathrm{i}}
\def\rmd{\mathrm{d}}

\makeatother
\setcitestyle{numbers,square}
\begin{document}

\title{Controlling symmetry and localization with an artificial gauge field in a disordered quantum system}

\author{Clément Hainaut}
\author{Isam Manai}
\author{Jean-François Clément}
\author{Jean Claude Garreau}
\author{Pascal Szriftgiser}
\affiliation{Université de Lille, CNRS, UMR 8523, Laboratoire de Physique des Lasers Atomes et Molécules, F-59000 Lille, France}

\author{Gabriel Lemarié}
\affiliation{Laboratoire de Physique Théorique, IRSAMC, Université de Toulouse, CNRS, 31062 Toulouse, France}
\affiliation{Department of Physics, Sapienza University of Rome, P.le A. Moro 2, 00185 Rome, Italy}

\author{Nicolas Cherroret}
\author{Dominique Delande}
\affiliation{Laboratoire Kastler Brossel, UPMC-Sorbonne Universités, CNRS, ENS-PSL Research University, Collège de France, 4 Place Jussieu, 75005 Paris, France}

\author{Radu Chicireanu}
\email{radu.chicireanu@univ-lille1.fr}
\affiliation{Université de Lille, CNRS, UMR 8523, Laboratoire de Physique des Lasers Atomes et Molécules, F-59000 Lille, France}

\maketitle

\textbf{Anderson localization, the absence of diffusion in disordered media, draws its origins from the destructive interference between multiple scattering paths. The localization properties of disordered systems are expected to be dramatically sensitive to their symmetry characteristics. So far however, this question has been little explored experimentally. Here, we investigate the realization of an artificial gauge field in a synthetic (temporal) dimension of a disordered, periodically-driven (Floquet) quantum system. Tuning the strength of this gauge field allows us to control the time-reversal symmetry properties of the system, which we probe through the experimental observation of three symmetry-sensitive  `smoking-gun' signatures of localization. The first two are the coherent backscattering, marker of weak localization, and the coherent forward scattering, genuine interferential signature of Anderson localization, observed here for the first time. The third is the direct measurement of the $\beta(g)$ scaling function in two different symmetry classes, allowing to demonstrate its universality and the one-parameter scaling hypothesis.}

\subsection*{Introduction}

Symmetry, disorder and chaos are ubiquitous in both classical and quantum physics. These concepts are intimately intertwined: In a `disordered crystal' for instance, disorder stems from the absence of translational symmetry. But this does not mean that symmetries are absent in disordered/chaotic systems; on the contrary, they play a central role as systems presenting the same symmetries display analogous properties. This idea led to the  fundamental concept of `universality class', grounding the famed random matrix theory~\cite{Haake:QuantumSigChaos:01}.  In a dirty metal for instance, breaking the `time-reversal symmetry' ($T$-symmetry) has a profound effect on transport observables like electrical and thermal conductivities~\cite{MaillySanquerUCF}. A popular way to break the $T$-symmetry for \emph{charged} particles is to add a magnetic field. For neutral systems, where magnetic fields are inoperative, the concept of `artificial gauge field'~\cite{Lin:SyntheticMagneticFieldsForUltracold:N09,LinBECinLightVectorPot,Lin2011SyntElectricField} has recently been introduced: It consists in building Hamiltonians which behave as if a gauge field were present. In the present work, we exploit the simplicity and flexibility of \emph{driven} cold-atom systems to generate such an artificial gauge field.  For this purpose, we build on the well-known \emph{kicked rotor}~\cite{Moore:AtomOpticsRealizationQKR:PRL95}, which has the status of a paradigm of both classical and quantum Hamiltonian chaos and can be mapped onto an Anderson-like Hamiltonian in \emph{any} dimension~\cite{Fishman:LocDynAnders:PRL82,Casati:IncommFreqsQKR:PRL89}. This system is realized experimentally by submitting laser-cooled atoms to `kicks' (constituting the driving) of a far-detuned laser standing wave.

By engineering the \emph{periodic} drivings, we obtain an experimental `knob' providing complete control of the relevant symmetry of the system, here the product of parity and time-reversal ($PT$-symmetry)~\cite{Blumel:SymmetryKR:PRL92,Blumel:SymmetryKR:PRE93,Scharf:KRForASpin1/2:JPAMG89}. Furthermore, we exploit the idea that the accumulated phase of a quantum particle along a closed multiple-scattering path is independent of the sense in which the loop is traveled when $PT$-invariance holds (defining the so-called `orthogonal class'), but not when it is broken (defining, for spinless systems, the `unitary class'), an effect that strongly affects quantum interference in localization phenomena. This allows us to directly observe the impact of this symmetry changing on interference signatures of localization in disordered media, and to study the universal transport properties in the two symmetry classes.

\subsection{Artificial gauge fields in disordered Floquet systems}

We first show how to engineer the driving of Floquet systems to manipulate their fundamental symmetry properties. For this purpose, we consider a generalized kicked rotor Hamiltonian, to which we add an amplitude $\mathcal{K}(t)$ and a spatial phase $a(t)$ in the potential term, both periodically modulated in time:
\begin{equation}
H=\frac{p^{2}}{2}+\mathcal{K}(t)\ \cos[x-a(t)]\ \sum_{n}{\delta(t-n)}\;,\label{eq:HKR}
\end{equation}
where $x$ and $p$ are the dimensionless position and momentum of the particle (see definitions in Appendix A). When $\mathcal{K}=\text{const.}$ and $a=0$, we recover the standard kicked rotor, which can be mapped onto an Anderson-like tight-binding model in momentum space~\cite{Fishman:LocDynAnders:PRL82,Moore:AtomOpticsRealizationQKR:PRL95} with on-site pseudo-disorder.

\begin{figure}
	\centering{}\includegraphics[width=16cm]{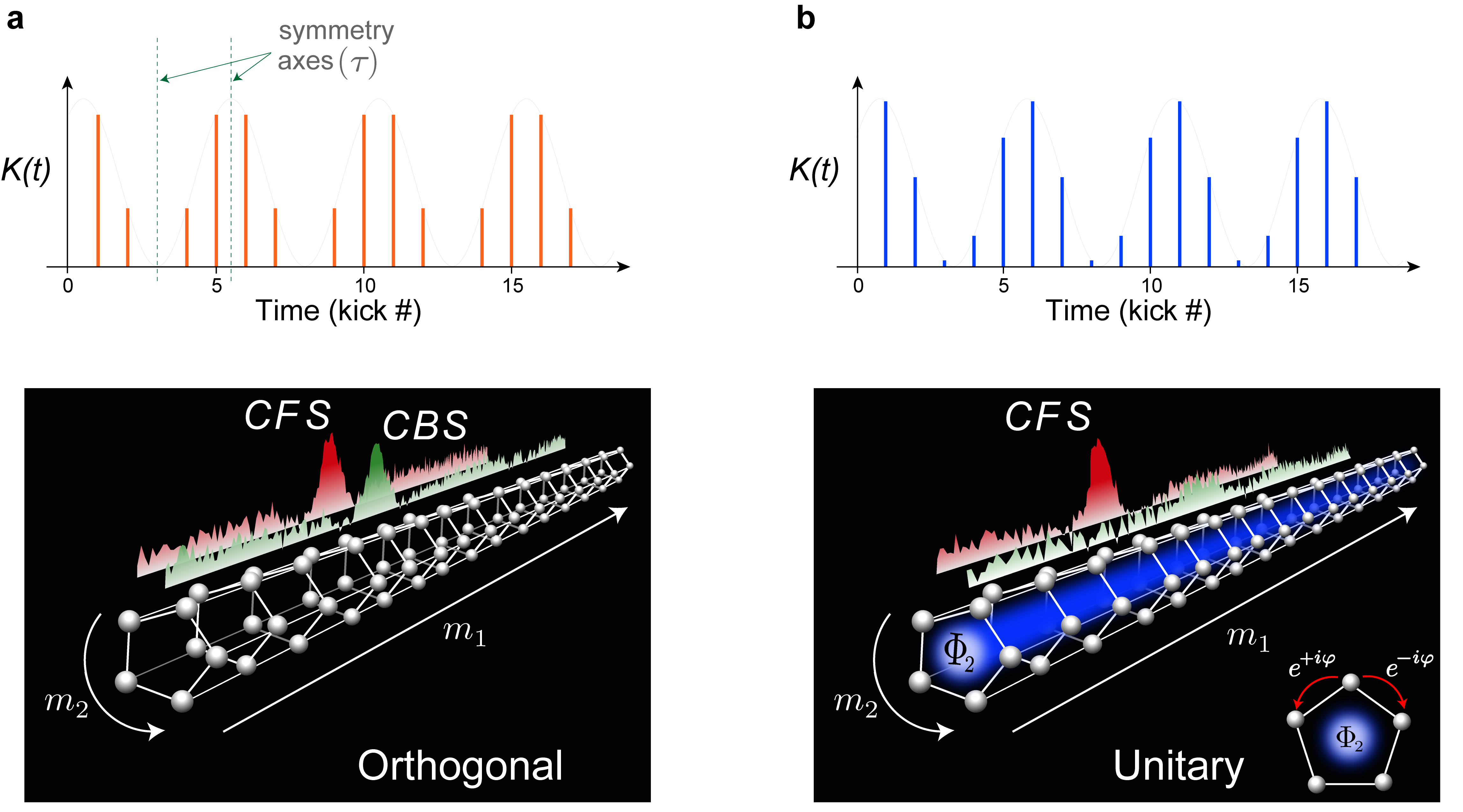} \caption{\label{fig:nanotube} \textbf{Emergence of an artificial gauge field in Floquet systems with periodically-modulated driving.} With a periodic kick amplitude (or phase) modulation, our system maps on a disordered `synthetic nanotube' in momentum space. By conveniently tailoring the temporal driving, we are able to create an artificial gauge field which controls the time-reversal symmetry properties. For a time-symmetric (\textbf{a}) kick sequence $\mathcal{K}(t)$ the system belongs to the orthogonal symmetry class, whereas a kick sequence without \textit{any} particular symmetry axes (\textbf{b}), corresponding to the presence of a non-zero Aharonov-Bohm flux $\Phi_{2}$ (sketched as the light blue area), puts the system in the unitary symmetry class (broken $T$-symmetry). Experimentally, two distinct interference signatures can be used to characterize symmetry and localization: the disappearance of the CBS peak is a clear-cut signature of the symmetry breaking, while the emergence of a CFS peak is a direct interference signature of the onset of Anderson localization, in both symmetry classes.}
\end{figure}

When $\mathcal{K}(t)$ is temporally modulated at a period $2\pi/\omega_2$ incommensurate with the kick period, it has been shown~\cite{Shepelyansky:Bicolor:PD87,Casati:IncommFreqsQKR:PRL89,Lemarie:AndersonLong:PRA09} that the temporal modulation can be taken into account by adding a `position' $x_{2}=\omega_{2}t+\varphi$ along a synthetic dimension labeled `2' (`1' refers to the physical dimension along which all measurements are performed). Here, we study the situation where the driving modulations have a period which is an \textit{integer} multiple of the kick period ($\omega_2=2 \pi /N$), i.e. $\mathcal{K}(t+N)=\mathcal{K}(t)$ and $a(t+N)=a(t)$ with $N$ an integer. In this case, the synthetic dimension is also periodic with \textit{twisted} boundary conditions. Such a system maps onto a synthetic \textit{nanotube} threaded by an artificial gauge field (see Fig.~\ref{fig:nanotube}). The flux of this artificial gauge field through the transverse section of the nanotube can be easily controlled by changing the initial phase $\varphi$ of the temporal modulation.

\bigskip{}

Without loss of generality, it is convenient to illustrate the fundamental mechanism of creation and control of the artificial gauge field by using the specific example of a period-$N$ amplitude modulation ($N=5$ in the experiment, see below):
\begin{equation}
\mathcal{K}(t)=K\left[1\!+\!\cos\left(\frac{2\pi t}{N}+\varphi\right)\right].\label{eq:kick-sequence}
\end{equation}
The temporal dynamics can be mapped on that of a two-dimensional pseudo-rotor with Hamiltonian~\cite{Lemarie:AndersonLong:PRA09}: $\mathcal{H}=p_1^{2}/2+2\pi p_{2}/N+K\cos x_{1}\left[1+\cos x_{2}\right]\ \sum_{n}\delta(t-n)$, where $x_{1}=x,p_1=p$ and the direction `2' is an ancillary space with $0\!\leq\!x_{2}\!<\!2\pi$, where the period $N$ dynamics is simply given by $x_{2}\!=\!\varphi+2\pi t/N\ (\mathrm{mod.}\ 2\pi)$. This equivalent 2D Hamiltonian is time-periodic with period 1. Its Floquet states \textendash{} eigenstates of the evolution operator over one period with eigenvalue $\rme^{\rmi\omega}$ \textendash{} are also solution of a tight-binding model: $\epsilon_{\mathbf{m}}\Psi_{\mathbf{m}}+\sum_{\mathbf{r}}W_{\mathbf{r}}\Psi_{\mathbf{m}-\mathbf{r}}=0$ where $\mathbf{m}\equiv(m_{1},m_{2})$ and $\mathbf{r}$ label the sites of a 2D square lattice which correspond to momenta in units of effective Planck's constant $\kbar$, and $\Psi_{\mathbf{m}}$ are the components of the Floquet quasi-states. The on-site energy $\epsilon_{\mathbf{m}}$ is $\epsilon_{\mathbf{m}}=\tan\left\lbrace \left[\omega-\left(\kbar{m_{1}}^{2}/2+2\pi m_{2}/N\right)\right]/2\right\rbrace $ and the hopping amplitudes $W_{\mathbf{r}}$ are coefficients of the twofold Fourier expansion of $W({x}_{1},{x}_{2})=\tan\left[K\cos{x}_{1}(1+\cos{x}_{2})\right/2\kbar]$~\cite{Casati:IncommFreqsQKR:PRL89}.

When $\kbar$ is incommensurate with $2\pi$, the on-site energies constitute a pseudo-random sequence in the direction `1', which accounts for the disordered character of our system \textit{in momentum space}, leading to Anderson localization. Nevertheless, the on-site energies are \textit{periodic} along the direction `2' with period $N$. Thus, we can use the Bloch theorem along the direction 2 and write any Floquet state as: $\Psi_{m_{1},m_{2}}=\rme^{-\rmi m_{2}\varphi_{2}}\ \psi_{m_{1},m_{2}}$, where $\varphi_{2}$ is the Bloch phase and $\psi_{m_{1},m_{2}+N}=\psi_{m_{1},m_{2}}$ is periodic in direction 2. Moreover, since the modulation phase is well-defined, the initial condition in direction 2 is simply $\delta(x_{2}-\varphi)=\sum_{m_{2}=-\infty}^{+\infty}{\rme^{-\rmi m_{2}\varphi}\rme^{\rmi m_{2}x_{2}}}$~\cite{Lemarie:AndersonLong:PRA09}, which selects the Bloch phase $\varphi_{2}\!=\!\varphi$.

The Hilbert space for $\Psi$ reduces thus to a synthetic nanotube along direction 1, with $N$ sites in the transverse section along direction 2, see Fig.~\ref{fig:nanotube}. The initial phase $\varphi$ of the temporal modulation controls the flux through the nanotube. Indeed, the Floquet eigenequation can be rewritten for the periodic
function $\psi$ as:
\begin{equation}
\epsilon_{m_{1},m_{2}}\psi_{m_{1},m_{2}}+\sum_{r_{1},r_{2}}W_{r_{1},r_{2}}\ \rme^{\rmi\varphi r_{2}}\ \psi_{m_{1}-r_{1},m_{2}-r_{2}}=0\label{eq:phase}
\end{equation}

The hopping matrix elements in Eq.~\eqref{eq:phase} have caught a phase $\varphi r_{2}$. This is similar to a 2D system exposed to a uniform magnetic field~\cite{KlitzingPRL1980IntegerQHE,HofstadterPRB1976}. However, the geometry here is not that of a planar system, but rather a quasi-1D system or a nanotube infinite along direction 1 and with $N$ transverse sites along direction 2. Indeed, a closed loop $m_{2}\!=\!0\!\to\!1\!\to\!2...\!\to\!N-1\!\to0$ will pick a total phase $\Phi_{2}=N\varphi$, while the counter-propagating loop will pick the opposite phase $-N\varphi.$ In contrast, no phase is picked along a plaquette $(m_{1},m_{2})\!\to\!(m_{1}+1,m_{2})\!\to\!(m_{1}+1,m_{2}+1)\!\to\!(m_{1},m_{2}+1)\!\to\!(m_{1},m_{2})$.
Thus, the effective gauge field flux $\Phi_2$ is similar to a magnetic flux, with the magnetic field along the axis `1' of the nanotube.

A generic value of $\varphi$ corresponds to a non-vanishing (mod.$\pi$) flux $\Phi_{2}$. In such a situation, it is not possible to unwind all the phases in Eq.~\eqref{eq:phase} so that the system is expected to be in the unitary symmetry class, where all anti-unitary symmetries \textendash{} product of time-reversal by a geometrical unitary operation \textendash{} are broken\footnote{The case $N=2$ is special, as the nanotube then degenerates in a two-leg ladder with a single transverse hopping matrix element. All phases can be unwound, and the system is the orthogonal class whatever $\varphi.$}. In contrast, if $\Phi_{2}=0$ (mod. $\pi$), all hopping terms can be made real and the system is expected to be in the orthogonal class:
\begin{eqnarray}
N\varphi & = & 0\ \ \mathrm{(mod.}\ \pi)\ :\ \mathrm{orthogonal\ class}\nonumber \\
N\varphi & \neq & 0\ \ \mathrm{(mod.}\ \pi)\ :\ \mathrm{unitary\ class.}\label{eq:condition}
\end{eqnarray}

This simple condition can also be deduced from a direct analysis of the kick sequence. For the kicked rotor~\eqref{eq:HKR}, the relevant anti-unitary symmetry is the product of time-reversal by parity ($PT$-symmetry)~\cite{Blumel:SymmetryKR:PRL92,Blumel:SymmetryKR:PRE93}. The Hamiltonian being explicitly time-dependent, there is not a single generalized time-reversal operator, but a family of operators $\mathcal{T}_{\tau}\!:\!t\!\to\!2\tau-t;\,x\!\to\!-x;\,p\!\to\!p$, depending on the temporal origin of the time reversal. The condition for $\mathcal{T}_{\tau}$ to be a symmetry operation requires that the sequence of kick amplitudes $\mathcal{K}(t)$ be symmetric around some time $\tau$ (Fig.~\ref{fig:nanotube},b). In the more general case of the Hamiltonian~\eqref{eq:HKR}, it requires additionally that the kick phases $a(t)$ be antisymmetric (as $\mathcal{T}_{\tau}$ changes $x$ to $-x$).

\subsection{Coherent Back and Forward Scattering}

Interference phenomena, which are at the core of Anderson localization, are very sensitive to symmetry breaking. Coherent Backscattering (CBS) is a simple example: a consequence of the $PT$-symmetry is that pairs of scattering paths associated with the same geometrical loop, but traveled in opposite senses, accumulate the same quantum phase and thus interfere constructively. When the symmetry is broken, these pairs of paths become out of phase and CBS disappears. However, in the presence of (strong) Anderson localization, other non-trivial quantum interference effects still exist, such as the Coherent Forward Scattering (CFS), recently predicted theoretically~\cite{Karpiuk:CFSFirst:PRL12} (see also~\cite{Lemarie16} in the context of the kicked rotor). Contrary to CBS, the CFS is present \textit{irrespective of the symmetry breaking} and, for unbound systems, \textit{requires the onset of Anderson localization} in order to show up~\cite{Ghosh:CFS2D:PRA14,Lee:CFS1D:PRA14,Micklitz:CFS1D:PRL14,Ghosh:CFS3D:PRA17}. While experimental observations of CBS have been achieved in many different systems (e.g.~\cite{Wolf:WeakLocCBSLight:PRL85,Bayer_93,Wiersma_95,Tourin_97,Jendrzejewski:CBSUltracoldAtoms:PRL12,Labeyrie:BSE:EPL12}), no observation of the CFS had been reported so far; here, we provide its first experimental observation.

In spatially-disordered systems, CBS and CFS manifest themselves in the reciprocal space as two peaks centered around $-\mathbf{k}^{0}$ (backward) and $+\mathbf{k}^{0}$ (forward direction, resp.) of the velocity distribution of a wave packet initially launched with a well-defined wave vector $\mathbf{k}^{0}$~\cite{Karpiuk:CFSFirst:PRL12}. Alternatively, the constructive interference between time-reversed loops manifests itself in the direct (configuration) space by an enhanced probability to return to the original position~\cite{AkkermansMontambaux:MesoscopicPhysics:11}.

This interference is visible, in our system, in a mixed momentum/configuration space representation $(p_1,x_2)$, in which the initial state is localized. Starting from $p_1(t=0)\approx0$ and $x_{2}(t=0)=+\varphi,$ a CBS peak should be observed around $p_1=0$ at $x_{2}=-\varphi$ (in the presence of the $PT$-symmetry) and a CFS peak around $p_1=0$ at $x_{2}=+\varphi$~\cite{Lemarie16}. Because of the time-dependence of $x_{2}(t)=x_{2}(0)+2\pi t/N$, we thus expect to observe CBS and CFS at \textit{different times}, depending on the initial condition $x_{2}(0)$ (see Appendix C). Both CBS and CFS are measurable in the physical dimension $p_1$ as peaks around the initial momentum $p_1\approx0.$ The temporal modulation is thus essential to separate them, so that they appear at different moments during the kick sequence\footnote{For the `standard' kicked rotor, which belongs to the orthogonal symmetry class, the CBS and CFS should exist \emph{simultaneously} as peaks centered around $p_1=0$. Their distinct experimental observation would be particularly challenging, in presence of limiting factors such as decoherence and finite-width initial state (see below).}.

\begin{figure}
	\centering{}\includegraphics[width=11.4cm]{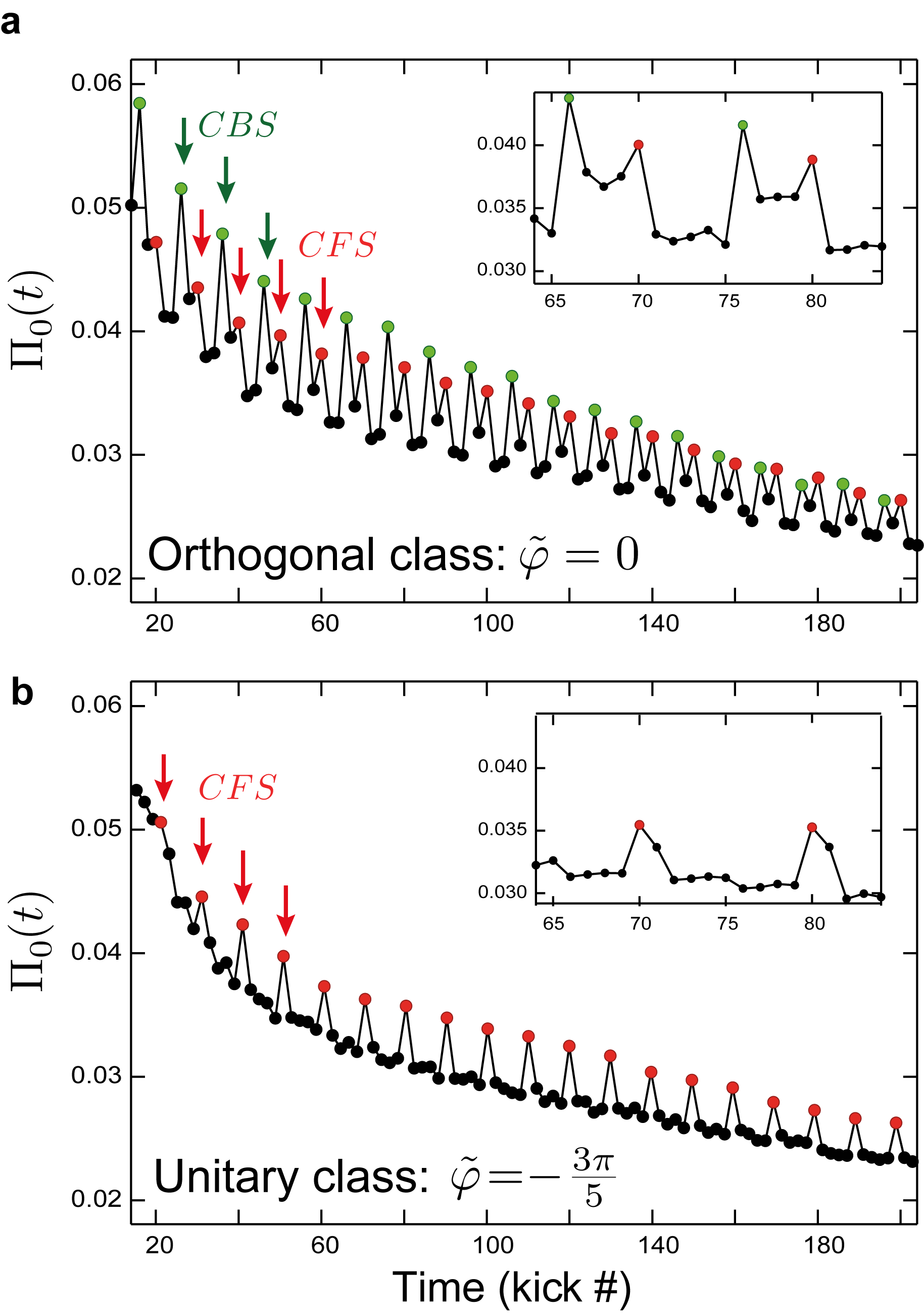}
	\caption{\label{fig:Pi0}  \textbf{Experimental observation of CBS and CFS peaks in two symmetry classes.} We measure the time-evolution of the zero-momentum probability density $\Pi_{0}(t)$ using the Hamiltonian~\eqref{eq:modulation_exp}, with the parameter $\tilde{\varphi}$ controlling the $PT$-symmetry properties. (\textit{Top}) In the orthogonal class ($\tilde{\varphi}=0$), we observe distinct enhancements of $\Pi_{0}(t)$ at times $t=6$ (mod.10) and $t=0$ (mod.10), associated to CBS (green) and CFS (red) peaks, respectively.  The CBS peaks have maximum contrast early during the kick sequence, and decrease due to stray decoherence, whereas the CFS peaks start by slowly increasing in contrast, and equalize the CBS at longer times. This constitutes a genuine interferential signature of the emergence of Anderson localization. (\textit{Bottom}) The time evolution of $\Pi_{0}$ obtained with a Hamiltonian with broken $PT$-symmetry ($\tilde{\varphi}=-3\pi/5$) clearly shows the disappearance of the CBS peaks in the unitary class. The CFS peaks, insensitive to the symmetry breaking, continue to be present, with a contrast following the same increasing trend at short times.}
\end{figure}

\bigskip{}

We experimentally studied the CBS and CFS effects by using a thermal, ultra-cold cloud of Cs atoms `kicked' by a series of short pulses of a far-detuned standing wave. It is created by a pair of counter-propagating laser beams, whose amplitude and relative phase can be changed from one kick to another in order to create any arbitrary sequences $\mathcal{K}(t)$ and $a(t)$. We measure, through time-of-flight, the `return probability', i.e. the zero-momentum probability density, $\Pi_{0}(t)=|\Psi(p_1 =0,t)|^{2}$ vs. time $t$. The flexibility of our system~\eqref{eq:HKR} allows us to optimize the properties of the experimental Hamiltonian (see details in Appendix B). We utilize a period-10 Hamiltonian, with a suitable combination of amplitude and spatial phase modulations: 
\begin{eqnarray}
\mathcal{K}(t) & = & K\left[1+\cos\left(\frac{2\pi(t-1)}{5}\right)\right]\ \ \mathrm{and}\ \ a(t)=-a,\ \ t\ \mathrm{even}\nonumber \\
\mathcal{K}(t) & = & K\left[1+\cos\left(\frac{2\pi(t-1)}{5}+\tilde{\varphi}\right)\right]\ \ \mathrm{and}\ \ a(t)=a,\ \ t\ \mathrm{odd.}\label{eq:modulation_exp}
\end{eqnarray}
The symmetry properties of the Hamiltonian are controlled by tuning the parameter $\tilde{\varphi}$, while the additional phase modulation $a(t)$, with period $2$, makes CBS and CFS observable only at even kicks (see Appendix C). The CFS peak is observed each time $x_{2}$ returns to its initial value, that is at kicks 10, 20, 30... The CBS peak is observed only if the Hamiltonian is $PT$-symmetric (amplitude-symmetric and phase-antisymmetric sequence). This is possible only if $\tilde{\varphi}$ is an integer multiple of $2\pi/5$. For example, for $\tilde{\varphi}=0$, the CBS peak is predicted to appear at kicks 6, 16, 26...

\begin{figure}
	\centering{}\includegraphics[width=12cm]{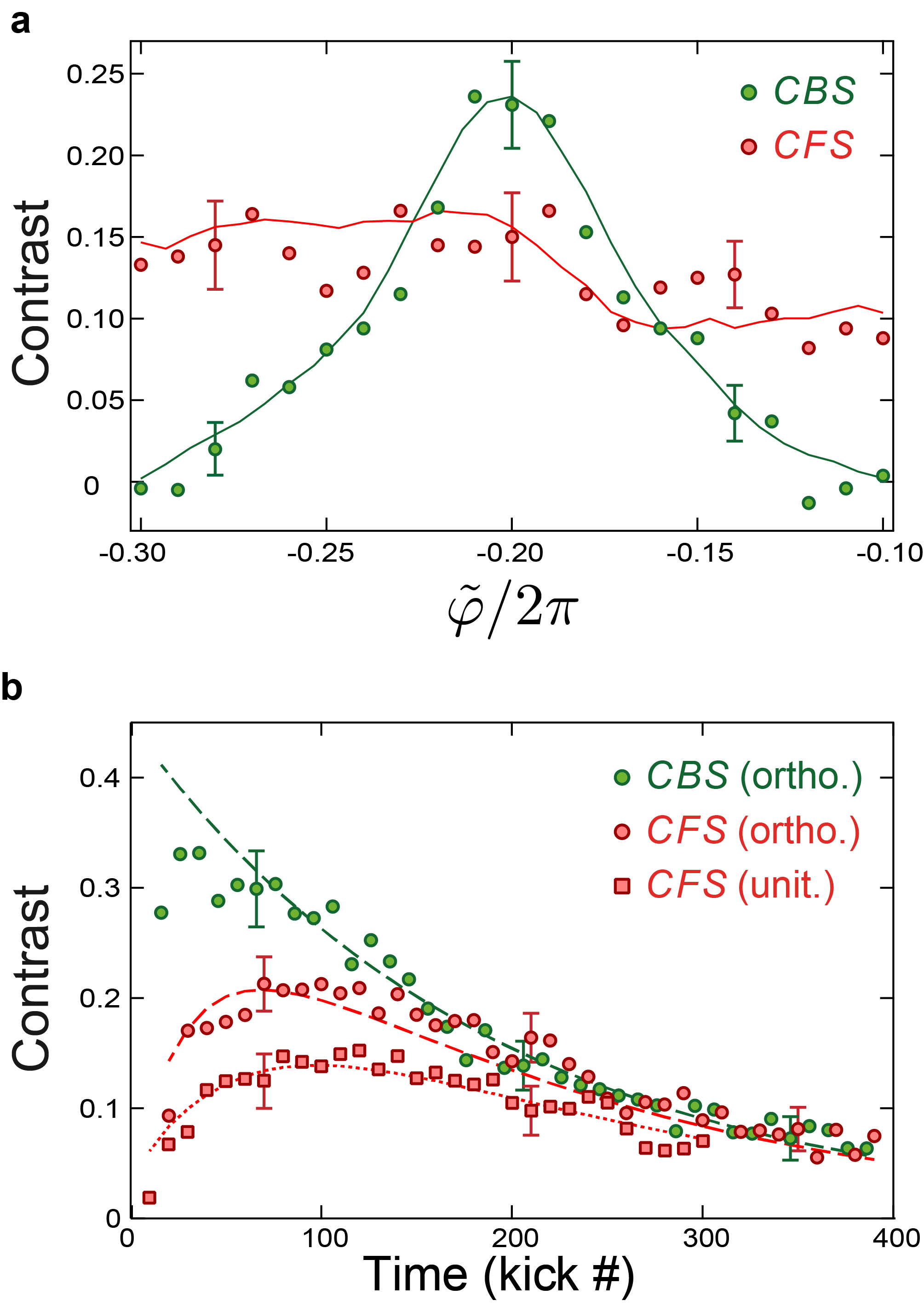} \caption{\label{fig:contrast_vs_b_time} \textbf{Temporal-dynamics and symmetry-breaking characteristics of the CBS and CFS peaks.} \textbf{a,} The experimental CBS (green) and CFS (red) contrasts were measured vs. the parameter $\tilde{\varphi}$, which controls the artificial gauge flux. The data is taken at $t=70$ kicks, when the CFS contrast approaches that of the CBS. The CBS contrast is maximum at $\tilde{\varphi}=-2\pi/5$, where there is a perfect $PT$-symmetry. When $\tilde{\varphi}$ varies, the CBS contrast decreases, and eventually vanishes when the symmetry is completely broken. In contrast, the CFS contrast is almost insensitive to the value of $\tilde{\varphi}$. The solid lines are ab initio numerical simulations using experimentally-measured parameters. \textbf{b,} The time evolutions of the CBS (orthogonal, green) and CFS (orthogonal -- red circles, and unitary -- red squares) contrasts corresponding to Fig.~\ref{fig:Pi0}. The CBS follows an exponential decay (dashed green line, fit), due to decoherence, with a fitted time constant $t_{\mathrm{dec}}\approx190$. The CFS contrasts are fitted using the equations in~\cite{Micklitz:CFS1D:PRL14} (red lines) with decoherence effects included. This yields $t_{\mathrm{loc}}\approx40$ in the unitary class and $t_{\mathrm{loc}}\approx16$ in the orthogonal class.}
\end{figure}

\bigskip

The experimental results for $\tilde{\varphi}=0$ (Fig.~\ref{fig:Pi0}a) display two characteristic features: First, the general trend of $\Pi_{0}(t)$ is a decay vs. $t,$ due to the spreading of the initially-narrow wave packet in momentum space. This decay slows down at long times, when localization sets in. Second, we observe pronounced peaks at kicks 20, 26, 30, 36, etc. From this series of peaks, one can however distinguish two subsequences with different properties: The CBS series at $t=6$ (mod.10) has a maximal contrast at the beginning, which slowly decreases with time, while the contrast of the CFS series at $t=0$ (mod.10) \textit{increases} at short times. On a longer time scale (set by the localization time $t_{\mathrm{loc}}$), the CFS amplitude asymptotically converges towards the CBS one, and the two peaks become twins after localization has set in. This constitutes a \textit{direct interferential proof} of the occurrence of Anderson localization.

Adding a phase $\tilde{\varphi}$ to the modulation creates an artificial gauge field which breaks the $PT$-symmetry. This directly manifests (Fig.~\ref{fig:Pi0}b) in the disappearance of the CBS peaks at $t=6$ (mod.10), whereas at pulses $t=0$ (mod.10) the CFS peaks survive and follow the increasing trend, until saturating at $t\sim t_{\mathrm{loc}}$.

To test their dependence on the artificial gauge field amplitude, we vary $\tilde{\varphi}$ and monitor the contrasts of the CBS and CFS peaks (see Appendix D for contrast definitions and measurement procedure). The results are shown in Fig.~\ref{fig:contrast_vs_b_time},a:
at $\tilde{\varphi}=-2\pi/5$ (which preserves the $PT$-symmetry) we observe a pronounced maximum of contrast for the CBS peaks, present here at kicks 2 (mod.10) (see Appendix C). The decrease of the CBS contrast around this value is a clear signature of the symmetry breaking. It is qualitatively similar to the magneto-resistance effect~\cite{bergman1984weak} induced in a solid-state sample when time-reversal symmetry is broken by an external magnetic field. On the other hand, the contrast of the CFS peak is insensitive to $\tilde{\varphi}$, showing its robustness vs. the $PT$-symmetry breaking.

There are fundamental differences between CBS and CFS dynamics: unlike the CBS peak, which is present at short times with maximal contrast, the CFS peak \textit{requires} (strong) Anderson localization in order to show up, on a time scale set by the localization time $t_{\mathrm{loc}}$. The time-dynamics of the CFS contrast has been theoretically predicted in~\cite{Micklitz:CFS1D:PRL14}, using a non-perturbative, fully time-resolved analytical description of a quantum quench in an Anderson-localized unitary system.

In our experiment (Fig.~\ref{fig:contrast_vs_b_time},b), the slow decay of both peaks at longer times is due to stray decoherence. The CBS contrast follows an exponential decay $C_{B}(t)=C_{0}\exp(-t/t_{\mathrm{dec}})$~\cite{Hainaut:ERO:PRL17} and is an excellent measure for the decoherence time $t_{\mathrm{dec}}$ on our system. A fit gives $t_{\mathrm{dec}}\approx190$ and an initial amplitude of the CBS contrast $C_{0}\approx0.45$  (which is lower than unity, due to a finite initial momentum width effect). In the unitary class, the CFS dynamics is very well fitted by the analytical formula of~\cite{Micklitz:CFS1D:PRL14} multiplied by the same exponential decay due to decoherence: $C_{F}(t)=C_{0}I_{0}(2t_{\mathrm{loc}}/t)\exp(-2t_{\mathrm{loc}}/t)\exp(-t/t_{\mathrm{dec}})$ with $t_{\mathrm{loc}}\approx40$ the only fitting parameter ($I_0$ is the modified Bessel function of order zero). The same fit can be applied in the orthogonal class (see~\cite{Lee:CFS1D:PRA14} for a numerical study) and gives a good agreement. As expected, we measure a smaller value of $t_{\mathrm{loc}}\approx16$, which is reduced because of the presence of `simple' loops favoring localization on a shorter time scale.

\bigskip{}

These observations prove that the CFS is a marker of non-trivial quantum interference needed to build Anderson localization in quantum disordered systems. The fact that we can observe a \textit{destruction} of CBS \textit{in the presence of a surviving CFS} is a clear-cut proof of the $PT$-symmetry breaking, and that other effects, such as decoherence, are not at the stake for the destruction of the CBS (Appendix B). Hence, this represents an unambiguous evidence of the changing of our system from the orthogonal to the unitary class under the effect of the artificial gauge field.

\subsection{Symmetry and transport: universal one-parameter scaling law}

The interference phenomena leading to Anderson localization also dramatically influence the transport behavior in the bulk of disordered quantum systems. First corrections to the `classical' (incoherent) diffusion coefficient $D_0$, known as weak localization, come from CBS-type interference which enhance the return probability of a quantum particle~\cite{AkkermansMontambaux:MesoscopicPhysics:11}. This quantum corrections are directly linked to the presence of the $PT$-symmetry. In absence of this symmetry, more complex CFS-type interference induce a smaller deviation from diffusive behavior, with a distinct form. 

As we will show below, the temporal evolution of the average momentum spread $\langle p_1^{2}(t)\rangle$ provides an excellent insight of the manner this happens in our system~\eqref{eq:HKR}. We measure $\langle p_1^{2}(t)\rangle$ with $\mathcal{K}(t)=K$ and $a(t)=a(t+N)$ a periodic series of $N$ randomly-chosen phases i.i.d. in $[0,2\pi]$, which in general breaks the $PT$-symmetry. However, the $PT$-symmetry can be restored by imposing the $a(t)$ series to be antisymmetric. The experimental results are averaged over a large number (100) of realizations of these random phases, with the microscopic parameters $K$, $N$ and $\kbar$ fixed, thus allowing both the very high precision necessary for studying the scaling properties and to vary the microscopic parameters to test the universality of the experimental observations (see below).

\begin{figure}
	\centering{}\includegraphics[width=14cm]{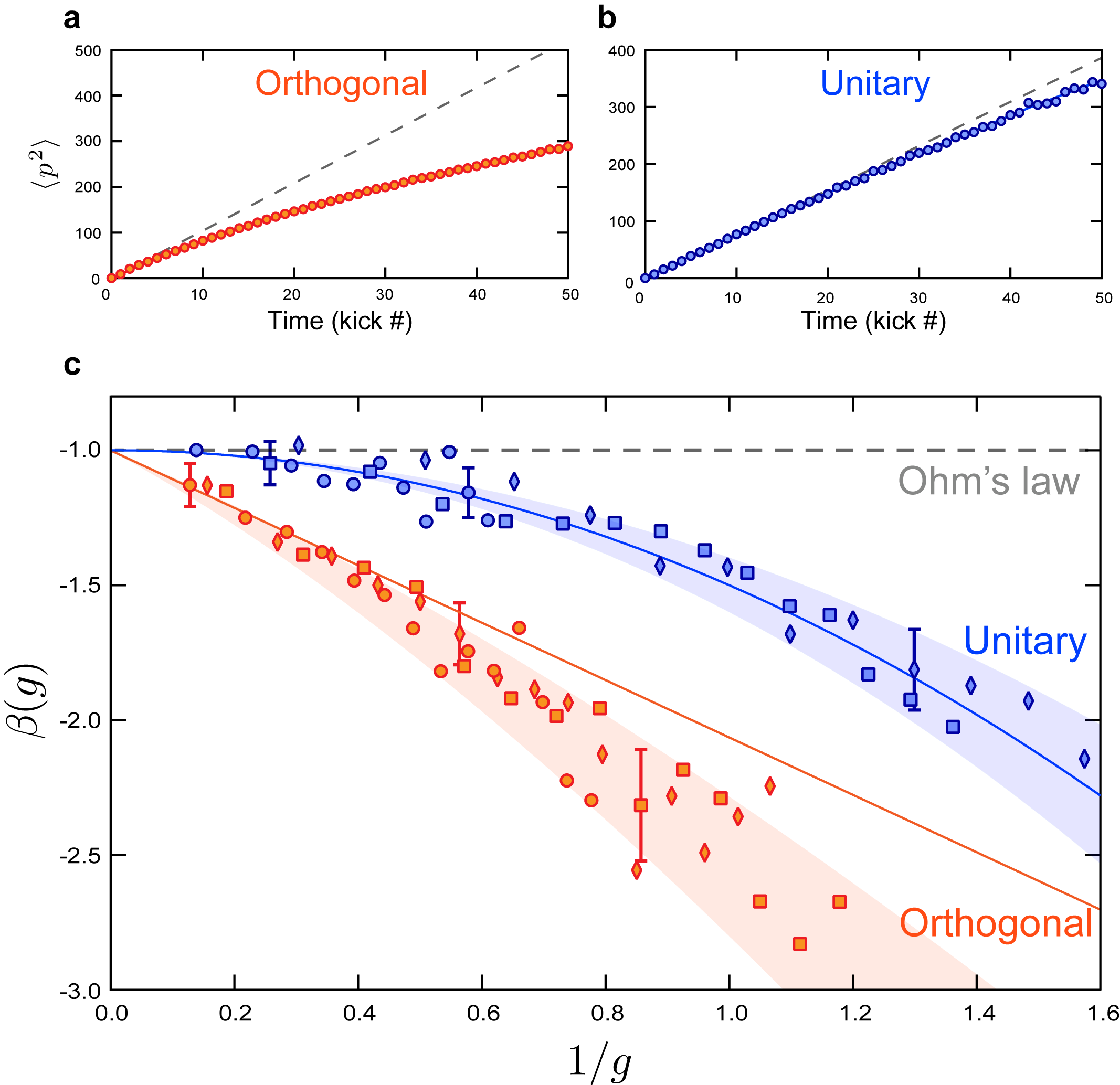} \caption{\label{fig:betaG} \textbf{Weak localization corrections and one-parameter scaling function $\beta(g)$ in a quasi-1D disordered system.} \textbf{a,b,} Time-evolution of $\langle p_1^{2}\rangle$ in the weak-localization regime in the two symmetry classes. Closed-loop corrections lead to a rapid deviation from classical diffusion (dashed line) in the orthogonal class (\textbf{a}). In the unitary class (\textbf{b}), these corrections are absent, which qualitatively translates in a much slower departure from classical diffusion. In both cases, $D_{0}$ is the same within $\sim20\%$. \textbf{c,} Experimental dependence of the $\beta(g)$ function on the dimensionless conductance $g= N\sqrt{\langle p_1^{2}\rangle}/(\kbar t)$, measured following the 1D spreading of a wave packet in momentum space. The different symbols (circles, diamonds and squares) correspond to three sets of different microscopic parameters ($K$ and $N$) of the system: $(K,N)\in\{(4,3),(4.5,4),(3.5,5)\}$ (orthogonal, orange) and respectively $(K,N)\in\{(2.5,3),(4,4),(1.6,5)\}$ (unitary, blue), for a value of $\kbar=1$. All data in each class collapse onto two distinct universal $\beta(g)$ functions, characteristic of each symmetry class, indicated by the shaded regions. The asymptotic behavior at large $g$ is correctly predicted by eqs.~\eqref{eq:beta} (continuous lines) inside their domain of validity.}
\end{figure}

In the absence of quantum interference, $\langle p_1^{2}\rangle$ evolves diffusively with time: $\langle p_1^{2}\rangle=2D_{0}t$. In the orthogonal class, self-intersecting (CBS-like) single-loop interference paths, which are already present from very short times, lead to a rapid deviation from classical diffusion (Fig.~\ref{fig:betaG}.a). In the unitary class, where the one-loop corrections are absent, this has a dramatic effect on transport properties, leading to a `slower' deviation from classical diffusion (Fig.~\ref{fig:betaG}.b).

\bigskip{}

An instrumental progress in the theory of metal-insulator transitions was the so-called `one-parameter scaling theory' introduced by Abrahams\emph{ et al}.~\cite{Abrahams:Scaling:PRL79}. It shows that, irrespective of the microscopic details of the system, transport properties should obey a universal scaling behavior, characterized by a single quantity, $\beta\equiv\rmd\ln g/\rmd\ln L$, the logarithmic derivative of the dimensionless conductivity $g$ with respect to the size $L$ of the system, which is a measure of transport. Expressed only as a function of the conductivity $g$ itself, the resulting $\beta(g)$ function is `universal', that is, independent of microscopic details. This function has played a central role in the study of disordered systems. Here we present a direct experimental measurement of the $\beta(g)$ scaling function, for both orthogonal and unitary class, and a test of its universality within each symmetry class. This approach can be directly tested in our case, in a remarkably simple manner, by studying the momentum spreading of a wave packet in one dimension. 

The kicked rotor, which maps on a pseudo-random Anderson model (see above), should also obey a one-parameter law. It is however a \textbf{dynamical} system, so that one has to build dynamical quantities -- which are the equivalent of the system size $L$ and the dimensionless conductance $g$. The natural choice for the system size $L$ is the number of lattice sites effectively populated. In momentum space, the lattice sites are momentum eigenstates separated by $\Delta p=\kbar$, so that we can define $L=\sqrt{\langle p_1^{2}(t)\rangle}/\kbar.$ Following~\cite{Cherroret:AndersonNonlinearInteractions:PRL14}, we can define for the standard kicked rotor $g=\sqrt{\langle p_1^{2}(t)\rangle}/t\kbar.$ In the classical regime where the dynamics is diffusive, this leads to $g=2D_{0}/\kbar L,$ a perfectly sensible result with a conductance decreasing like the inverse of the system size (Ohm's law) and proportional to the diffusion coefficient (Einstein law). Such a definition immediately leads to $\beta=-1,$ as expected for a classical diffusive one-dimensional system.

However, one needs to take into account the fact that our synthetic quasi-1D system consists of $N$ transverse channels (see Fig.~\ref{fig:nanotube}). In this case, its conductance is $N$ times larger than for a purely-1D system. We thus define $g \equiv N\sqrt{\langle p_1^{2}(t)\rangle}/(t\kbar),$ while the definition of $L,$ counting the number of occupied lattice sites in the longitudinal direction `1' is unaffected.

The leading corrections to $\langle p_1^{2}(t)\rangle$ due to loops have been calculated for the kicked rotor in~\cite{Tian:EhrenfestTimeDynamicalLoc:PRB05}, both in the orthogonal and unitary classes\footnote{The correction in the unitary class is given with the wrong sign in~\cite{Tian:EhrenfestTimeDynamicalLoc:PRB05} (C. Tian, private communication).}. They allow us to compute the lowest-order correction to the $\beta(g)$ function, valid in the limit of large conductivities:
\begin{eqnarray}
\beta(g) & = & -1-\frac{4\sqrt{2}}{3\sqrt{\pi}g}\ :\ \mathrm{orthogonal\ class,} \nonumber \\
\beta(g) & = & -1-\frac{1}{2g^{2}}\  \;\;\;\; :\ \mathrm{unitary\ class.} \label{eq:beta}
\end{eqnarray}

In order to test these predictions, and the universality of $\beta(g)$, we studied a series of different values for the microscopic parameters $K$ and $N$, in the two symmetry classes. The measured $\beta$-functions are shown in Fig.~\ref{fig:betaG}.c. A remarkable feature of these results is that all data collapse on two distinct scaling functions, as evidenced by the shaded zones, characteristic for each universality class. This constitutes an experimental demonstration of the validity of the one-parameter scaling law. It also shows that the shape of the $\beta(g)$ function makes is a clear marker of the presence or absence of an artificial gauge field governing the $PT$-symmetry.

The unitary case is in excellent agreement with~\eqref{eq:beta}. This is also true in the orthogonal class, in the limit of large $g$ (typically for $1/g<0.5$). For smaller values of $g$ we notice deviations from~\eqref{eq:beta}, which we confirmed through numerical simulations. This probably indicates that higher-order interference diagrams should be taken into account in the orthogonal class, and we hope that these observations will stimulate further theoretical investigations in this direction.

\subsection{Conclusions et perspectives}

These striking observations highlight the importance of symmetries for the localization and transport properties of disordered media, and the possibility to control them using an artificial gauge field -- generated here by appropriately tailoring the driving parameters of a Floquet system. Our method presents a remarkable experimental simplicity, and avoids both the complexity and limitations in more involved schemes (using, e.g., close-to-resonance Raman-dressing of internal states). We characterized the Anderson localization from a new perspective, by directly probing interferential 'building blocks' such as the Coherent Back- and Forward- Scattering phenomena. We also measured, in perfectly controlled conditions, the $\beta(g)$ scaling function -- a universal characteristic measure of transport in disordered media. Moreover, we demonstrated the different sensitivity of these effects with respect to the artificial gauge field flux, which controls the $PT$-symmetry properties of the system.

Interference signatures (such as the CFS) could provide valuable tools to observe the Anderson transition and probe its critical properties in higher dimensions and different symmetry classes. Engineering spin-orbit-coupled dynamical Floquet systems (e.g. using internal-state-dependent optical potentials) would allow, for example, to study the symplectic symmetry class, where Anderson localization is expected to occur in dimensions as low as two. This also opens an avenue for the study of fascinating phenomena, like quantum Hall effect, Floquet topological insulators and artificial magnetism.

\subsubsection*{Acknowledgments}

The authors are grateful to Chushun Tian and Adam Rançon for fruitful discussions and to Antoine Browaeys for a critical reading of this manuscript. GL acknowledges an invited professorship at Sapienza University of Rome. GL thanks CalMiP for access to its supercomputer. This work is supported by Agence Nationale de la Recherche (Grant K-BEC No. ANR-13-BS04-0001-01), the Labex CEMPI (Grant No. ANR-11-LABX-0007-01), Programme Investissements d'Avenir under the program ANR-11-IDEX-0002-02, reference ANR-10-LABX-0037-NEXT, and the Ministry of Higher Education and Research, Hauts de France Council and European Regional Development Fund (ERDF) through the Contrat de Projets Etat-Region (CPER Photonics for Society, P4S).

\clearpage

\section*{Supplemental Material}

\subsection*{Appendix A: Experiment} \label{AppendixA}

In the experiment, we start from a laser-cooled Cesium atomic sample, prepared in a thermal state ($T \simeq1.5$ $\mu$K). The cloud `kicked' along the vertical $x$ axis by a far-detuned, pulsed (period $T_1$) optical 
standing wave (SW), which is created by two \textit{independent} lasers beams. This allows us to control the 
amplitude and phase of the potential (via the RF signal sent to two different AOMs) and to shape the  
modulation sequences $\mathcal{K}(t)$ and $a(t)$ as in~\eqref{eq:HKR}. The laser parameters are: the detuning $\Delta = -13$ GHz (at the Cs D2 line, wavelength $\lambda = 852.2$ nm), the maximum intensity $I=300$ mW/beam, the pulse duration $\tau=200$ ns, while $T_1$ is varied typically between $10$ and $30$ $\mu$s. After the desired number of kicks, the cloud is allowed to expand for $\sim170$ ms and the momentum distribution $|\Psi(p)|^{2}$ is measured using the time of flight technique. To determine $\langle p^2 (t)\rangle $, used for the $\beta(g)$ measurements, we fit the clouds' distribution of squared-momentum $|p \Psi(p)|^{2}$ using the Lobkis-Weaver formula~\cite{LobkisWeaver:SelfConsistentTransportLocWaves:PRE05}, which describes the dynamics of the wave function at all times, from the diffusive to the localized regime.

For the CBS/CFS measurements, it is crucial to utilize a sample with an initial momentum distribution narrower than the width of a Brillouin zone. Indeed, the CBS and CFS peaks have widths given by that of the initial state, and their respective contrasts (equal to one in the ideal case) is strongly reduced otherwise. In order to decrease the mean kinetic energy of the sample, the atoms are loaded in a very shallow 1D optical lattice (vertical direction), whose depth is less than the initial temperature. This filters out the most energetic atoms. Subsequently, we realize 1D adiabatic cooling by switching off the lattice in $\sim1\,\mu$s, reaching a momentum distribution width $< 0.67 \times 2\hbar k_L$, which corresponds to an `equivalent 1D temperature' $<400$ nK (this value is limited by the resolution of the time-of-flight detection).

\textbf{Units:} We have chosen conveniently-scaled variables in order to express the Hamiltonian in the dimensionless form~\eqref{eq:HKR}: distances along the $x$ axis are measured in units of $(2 k_L)^{-1}$ (where $k_L$ is the SW wave number), time in number of kicks (or units of $T_1$), the particle mass is unity. The Hamiltonian~\eqref{eq:HKR} is associated with the Schrödinger equation $i \kbar \frac{\partial\psi}{\partial t} =\hat H \psi$, where $\kbar \equiv 4 \hbar k_L^2 T_1/M$ plays the crucial role of an effective Planck constant, which can be adjusted at will by modifying e.g. the kick period $T_1$. The canonical commutation relation reads $[\hat x,\hat p]=i\kbar$.

The Hamiltonian~\eqref{eq:HKR} is spatially $2\pi$-periodic, so that the solutions of the Schrödinger equation can always be expanded on a discrete lattice in \textit{momentum space} $p_{m}=(m+\beta)\kbar$ where $\kbar$ denotes the effective Planck's constant, $m$ is an integer and $\-1/2<\beta\leq1/2$ is the Bloch vector varying in the first Brillouin zone. Due to the spatial periodicity of the system, $\beta$ is a constant of motion, so that the whole analysis can be performed for $\beta=0$.

\subsection*{Appendix B: Correlations and decoherence in the kicked rotor} \label{AppendixB}

Decoherence is a major limitation in both CBS/CFS and $\beta(g)$ experiments. In our experiment, it comes mainly from residual spontaneous emission and fluctuations in the SW phase. To keep decoherence under control, we use rather small \textit{average} values of $K$, where short-time correlations between kicks are known to occur (leading, for instance, to well-known oscillations in the diffusion coefficient~\cite{Rechester:Correl:PRL1980,Shepelyansky:Bicolor:PD87}). In our case, these temporal correlations are responsible for large-amplitude oscillations, affecting the measurements of both the CBS/CFS contrast, but \textit{especially} (via $\langle p^2 (t)\rangle $) of the $\beta(g)$ function. Indeed, because $\beta(g)$ is a logarithmic derivative ($\rmd\ln g/\rmd\ln L$), it is extremely sensitive to correlations, as well as to the experimental noise.

We are able to eliminate almost completely the effect of correlations by conveniently averaging over several realizations of the disorder. The best possibility is to average over a large number of realizations of the random phase sequence $a(t)$, this method being extensively used for our $\beta(g)$ measurements. Each experiment is repeated $~500$ times, with a total of $100$ different random realization of $a(t)$ (corresponding to as many different realization of disorder), and the resulting momentum distributions $|\Psi(p)|^{2}$ are averaged. While phase modulations are very convenient for averaging out the correlations, it turns out that using amplitude modulations $\mathcal{K}(t)$ and a relatively large ($\gtrsim 10$) modulation period is more suitable for achieving a proper temporal separation of the CBS and CFS peaks. In order to resolve the CFS dynamics, one also needs a sufficiently large $t_{\mathrm{loc}}$. For the kicked rotor this is usually achieved by increasing the kick amplitude $K$, which unfortunately decreases the decoherence time $t_{\mathrm{dec}}$ in the experiment. However, it turns out that adding a period-two phase modulation increases, for certain values of the phase-shift $a$, the diffusion coefficient $D_0$ (and thus $t_{\mathrm{loc}}$) without affecting $t_{\mathrm{dec}}$. For the experiments shown in Fig.~\ref{fig:Pi0} and~\ref{fig:contrast_vs_b_time}, a fixed value $a=0.21\times 2\pi$ was used.

This is why, for optimizing the experimental conditions for the measurements of the CBS and CFS contrast dynamics, we opted for a combination of phase and amplitude modulations~\eqref{eq:modulation_exp}.

\subsection*{Appendix C: Symmetry and times of occurrence of CBS and CFS peaks} \label{AppendixC}

\begin{figure}
	\centering{}\includegraphics[width=11cm]{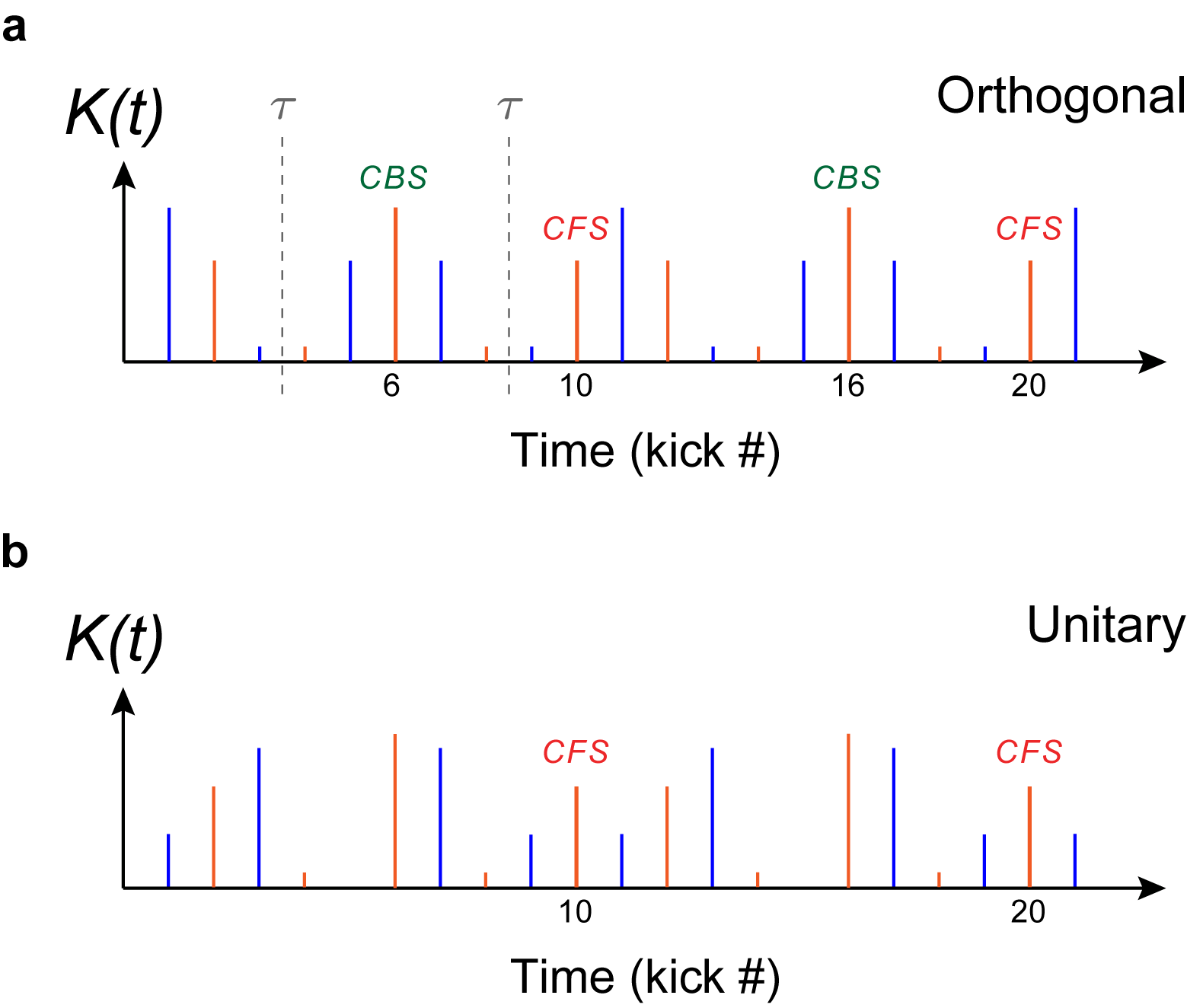}
	\caption{\label{fig:SeqExp} \textbf{Two pulse sequences belonging to different symmetry classes.} The sequences correspond to the data shown in Fig.~\ref{fig:Pi0}, and were obtained using two different values of the symmetry-control parameter $\tilde{\varphi}$ in~\eqref{eq:modulation_exp}: $\tilde{\varphi}=0$ (\textbf{a}) and $\tilde{\varphi}=-3\pi/5$ (\textbf{b}), for which the system belongs respectively to the orthogonal and unitary symmetry class. In the orthogonal class the time sequence has symmetry axes $\tau$; a CBS peak will appear at the kicks symmetric to the initial kick with respect to such axis. In the unitary class no CBS peak will exist. In both cases, the symmetry-insensitive CFS peaks are expected to occur at integer multiples of the period ($N=10$).}
\end{figure}

The pulse sequence is modulated using a combination of amplitude and phase modulations, as in~\eqref{eq:HKR}. The kick amplitude sequence $\mathcal{K}(t)$ has a period of 5, whereas the phase $a(t)$ is modulated with a period of 2 (represented in Fig.~\ref{fig:SeqExp} by the different colors used for the even and odd kicks), with an overall period $N=10$. A consequence of the period-two phase modulation $a(t)$ is that $PT$-symmetry axes only occur in-between kicks (and never \textit{during} a kick) which explains why CBS peaks do not occur for odd values of the kick number. A simple analysis of~\eqref{eq:modulation_exp} shows that the corresponding Hamiltonian is $PT$-symmetric (belonging thus to the orthogonal class) when the phase $\tilde{\varphi}\in2\pi\times\left\{0,\frac{1}{5},\frac{2}{5},\frac{3}{5},\frac{4}{5}\right\}$. Each of these values of the $\tilde{\varphi}$ leads to a different time of occurrence of the CBS peak -- corresponding to kicks $\left\{ 6, 10, 4, 8, 2 \right\}$ respectively.

Take for instance the modulation sequence shown in Fig.~\ref{fig:SeqExp}, corresponding to experimental data in Fig.~\ref{fig:Pi0}. When $\tilde{\varphi}=0$ (Fig.~\ref{fig:SeqExp},a), the sequence has $PT$-symmetry axes (vertical dashed lines labeled, $\tau$), and the system belongs to the orthogonal class. In this case, CBS peaks are expected to appear periodically, at kicks 6 (mod.10), i.e. at times equal to twice the occurrence time of each $\tau$. On the other hand, when $\tilde{\varphi}\notin 2\pi\times\left\{0,\frac{1}{5},\frac{2}{5},\frac{3}{5},\frac{4}{5}\right\}$ no symmetry axes exist (e.g. in (Fig.~\ref{fig:SeqExp},b), for $\tilde{\varphi} =-3\pi/5$). In both universality classes the symmetry-insensitive CFS peaks occur at integer multiples of the period of the system, i.e. at kicks 0 (mod.10).

\subsection*{Appendix D: CBS and CFS contrast measurements} \label{AppendixD}

Analyzing the experimental data in Fig.~\ref{fig:Pi0}, we can extract the contrasts $C_{B}(t)$ and $C_{F}(t)$, of the CBS and CFS peaks respectively, vs. time. The contrasts, for either case, are defined as: $C_{B,F}(t)=\left(\Pi_{0}(t)-\Pi_{0,\textrm{incoh.}}(t)\right)/\Pi_{0,\textrm{incoh.}}(t)$, and are evaluated at the occurrence times of their respective peaks, $t_{CBS}$ and $t_{CFS}$ (corresponding respectively to red and and green points in Fig.~\ref{fig:Pi0}). Here, $\Pi_{0}(t)=|\Psi(p_1 =0,t)|^{2}$ is the \textit{total} zero-momentum probability density (also defined in the main text), while $\Pi_{0,\textrm{incoh.}}(t)$ corresponds to the incoherent (or `classical') contribution to $\Pi_{0}(t)$. Outside $t_{CBS}$ and $t_{CFS}$ (i.e. at times corresponding to the black points in Fig.~\ref{fig:Pi0}), the two contributions are identical: $\Pi_{0}(t)=\Pi_{0,\textrm{incoh.}}(t)$. In order to evaluate $C_{B,F}(t)$, we interpolate the $\Pi_{0,\textrm{incoh.}}(t)$ values at $t_{CBS}$ and $t_{CFS}$. This method was used for the data shown in Fig.~\ref{fig:contrast_vs_b_time}.

\end{document}